\documentclass[final,5p,times,twocolumn]{elsarticle}

\usepackage{amssymb}
\usepackage[hidelinks]{hyperref}
\usepackage{color}
\hypersetup{breaklinks=true}

\makeatletter
\def\ps@pprintTitle{%
 \let\@oddhead\@empty
 \let\@evenhead\@empty
 \def\@oddfoot{}%
 \let\@evenfoot\@oddfoot}
\makeatother

\newcommand{\bbonu}{\ensuremath{\beta\beta0\nu}}

\newcommand{\XE}{\ensuremath{{}^{136}\rm Xe}}

\begin{document}

\begin{frontmatter}

\title{Helium-Xenon mixtures to improve the topological signature in high pressure gas xenon TPCs }

\author[label1]{R. Felkai}

\author[label2]{F. Monrabal\corref{mycorrespondingauthorprime}}
\cortext[mycorrespondingauthorprime]{Corresponding author}
\ead{francesc.monrabalcapilla@uta.edu}

\author[label3]{D. Gonz\'alez-D\'iaz}
\author[label1]{M. Sorel}
\author[label1,label2]{N. L\'opez-March}
\author[label1]{J.J.G\'omez-Cadenas}

\address[label1]{Instituto de F\'isica Corpuscular, CSIC \& Universitat de Val{\`e}ncia, Valencia, Spain}
\address[label2]{Department of Physics, University of Texas Arlington, USA}
\address[label3]{Instituto Galego de F\'isica de Altas Enerx\' ias (IGFAE), Universidade de Santiago de Compostela, Spain}

\begin{abstract}
Within the framework of xenon-based double beta decay experiments, we propose the possibility to improve the background rejection of an electroluminescent Time Projection Chamber (EL TPC) by reducing the diffusion of the drifting electrons while keeping nearly intact the energy resolution of a pure xenon EL TPC.
Based on state-of-the-art microscopic simulations, a substantial addition of helium, around 10 or 15~\%, may reduce drastically the transverse diffusion down to 2.5~mm/$\sqrt{\textnormal{m}}$ from the 10.5~mm/$\sqrt{\textnormal{m}}$ of pure xenon. The longitudinal diffusion remains around 4~mm/$\sqrt{\textnormal{m}}$.
Light production studies have been performed as well. They show that the relative variation in energy resolution introduced by such a change does not exceed a few percent, which leaves the energy resolution practically unchanged.
The technical caveats of using photomultipliers close to an helium atmosphere are also discussed in detail.
\end{abstract}


\end{frontmatter}


\renewcommand{\thetable}{\Alph{table}}

\section{Introduction}
\label{sec:intro}

Double beta decay is a process that has been observed for very few nuclei in its two-neutrino mode.
The unambigous observation of a neutrinoless double beta decay ($0\nu\beta\beta$) would definitely establish neutrinos as Majorana particles, which would ultimately demonstrate the existence of lepton number violating processes \cite{Schechter:1980gr}.
Luckily, one of these nuclei of interest is an isotope of Xenon, \XE . Being a noble gas, Xenon can be used in Time Projection Chambers (TPCs) where the target mass is actually the detection volume.
The scalability offered by xenon as a detector medium is a key factor in order to probe the entire inverted hierarchy, which would require reaching a maximum sensitivity to the effective Majorana mass of the electron neutrino, $m_{\beta\beta}$, of about 15~meV. In order to reach this sensitivity, ton scale detectors with less than 1~count/year in the region of interest are a must.
The grail of all rare event experiments, which is exacerbated in neutrinoless double beta experiments, is to reach the background free regime.

In this context, one of the specific technologies under developement is the electroluminescent (EL) high pressure xenon Time Projection Chamber \cite{Nygren:2009zz}, currently led by the NEXT collaboration \cite{Monrabal:2016chp} that is using a plane of 1~mm$^2$ SiPMs at 10~mm pitch and 8~mm from the center of the EL region to perform the tracking of the events. Several advantages of this design include good energy resolution, in the sub-percent range at Q$_{\beta\beta}$ (2459~keV), and the ability to perform topological reconstruction of events at this energy.

An $\sim$MeV electron moving through the gas loses its energy at a relatively constant rate until the end of its path where the energy deposition rate increases.
As a result, a fully contained ionization trail left by such an electron showcases a `blob-like' end-point.
The topology expected from a double beta event consists then of two electron tracks fully contained in the fiducial volume with a common origin and two `blobs' at their ends.
The main background source around Q$_{\beta\beta}$ stems from the $\gamma$-rays emitted from $^{208}$Tl and $^{214}$Bi events for which one of the end-points of the resulting track is misidentified as a blob. In addition, any characteristic X-rays emitted during the interaction of these gammas must convert relatively close to the main ionization track to avoid being clearly separated from it.
While the ionization trail is drifted toward the EL region, the diffusion of the ionization electrons degrades the imaging performance of the TPC.
Limiting its impact down to the level of the technical limitation set by the pixel pitch and EL gap thickness will allow for improved background rejection and therefore a reduced background rate.

Hence, the topological resolution is limited by instrumental factors, tracking plane segmentation and the width of the EL region, and physical limitations due to the diffusion of the drifting electron cloud. Diffusion is particularly large in pure xenon (see \cite{Renner:2016trj} for further discussion specific to NEXT detectors).
After one meter of drift, a point-like ionization deposit becomes a cloud distributed as a gaussian of 10~mm sigma in the direction perpendicular to the electric field (transverse) and 4~mm in the parallel direction (longitudinal).
This situation is far from ideal and can be largely improved by adding molecular electron coolants to the gas \cite{Azevedo:2017egv, Henriques:2017rlj} or by positive-ion detection \cite{Arazi:2017exp}.
As a reference, the thermal diffusion limit which can be found in \cite{Gonzalez-Diaz:2017gxo} gives a diffusion factor of $\sim$1.5~mm/$\sqrt{\textnormal{m}}$ for a field of 250~V/cm, which is very close to the $\sim$2.5~mm/$\sqrt{\textnormal{m}}$ value obtained for instance in Xe/CO$_2$ mixtures \cite{Henriques:2017rlj}.

Looking forward, this paper explores the possibility of using a substantial addition of helium to reduce the transverse diffusion while keeping the energy resolution intact. Helium is a particularly interesting alternative to the use of molecular additives, being easier to handle and expectedly free from light quenching effects. Its working principle and main enabling assets are sketched in this communication.
\section{Electron cooling and diffusion}
\label{sec:ecool}

The high diffusion of electrons drifting in heavy noble gases is a well known issue.
For VUV-quenched gas mixtures, as those commonly used in the operation of gaseous detectors, the presence of molecular additives (CH$_4$, CO$_2$, C$_4$H$_{10}$, ...) can be used advantageously in order to adjust diffusion.
The low energy rotational and vibrational states of these molecules allow the electrons to cool down very effectively leading to very low diffusions.
This solution, applied to an EL Xenon TPC, is detailed in \cite{Azevedo:2015eok}.
In this section we discuss the diffusion in pure noble gases and explain the mechanisms by which adding helium significantly reduces diffusion in xenon.
We also present results of simulations demonstrating that helium-doped xenon is a serious candidate in the prospect of lowering the gas diffusion, maintaining the energy resolution of pure xenon at the same time.

\subsection{Transverse diffusion}
\label{subsec:transdiff}

\begin{figure}[ht!]
	\includegraphics[width=0.48\textwidth]{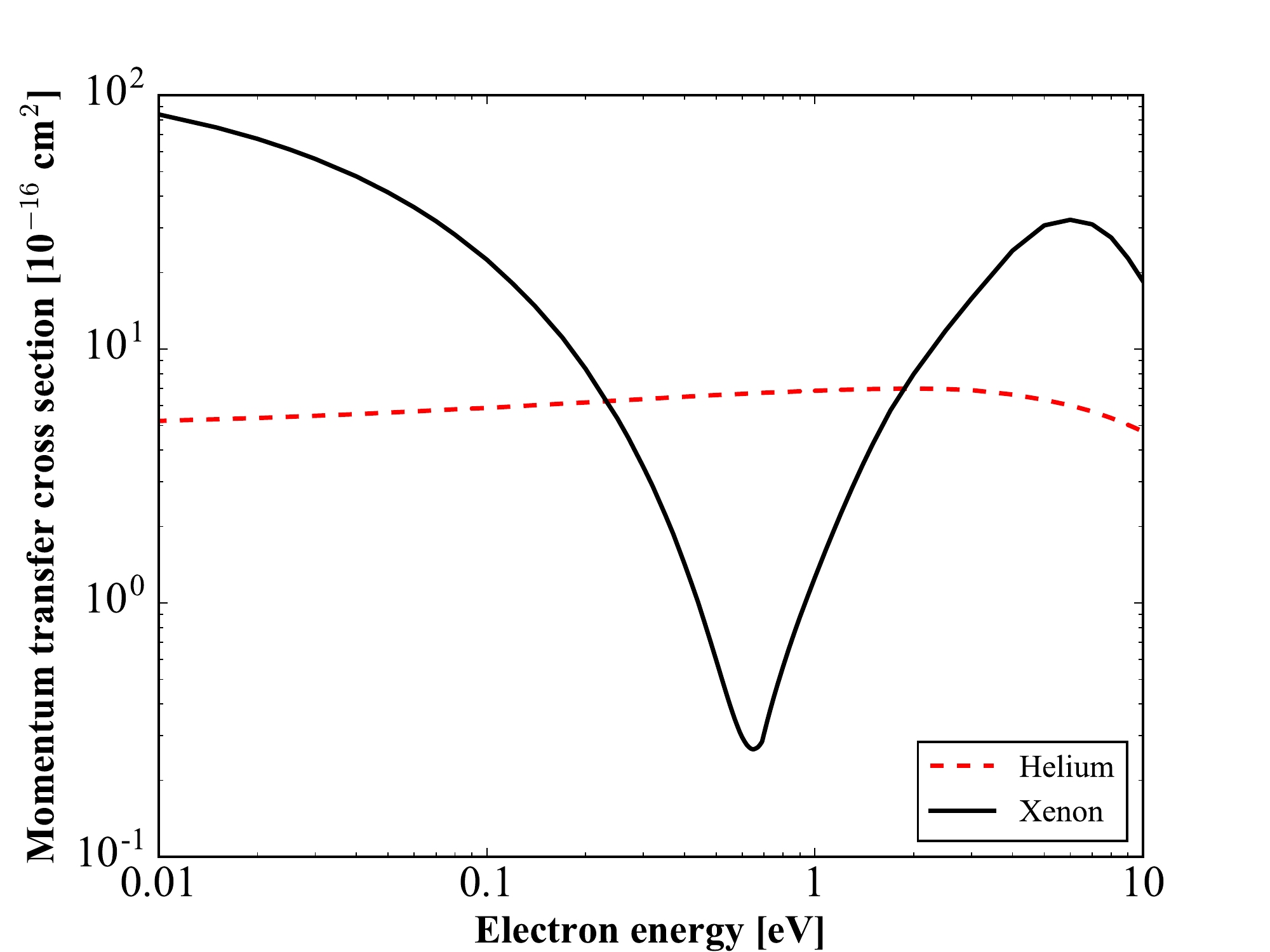}
	\includegraphics[width=0.48\textwidth]{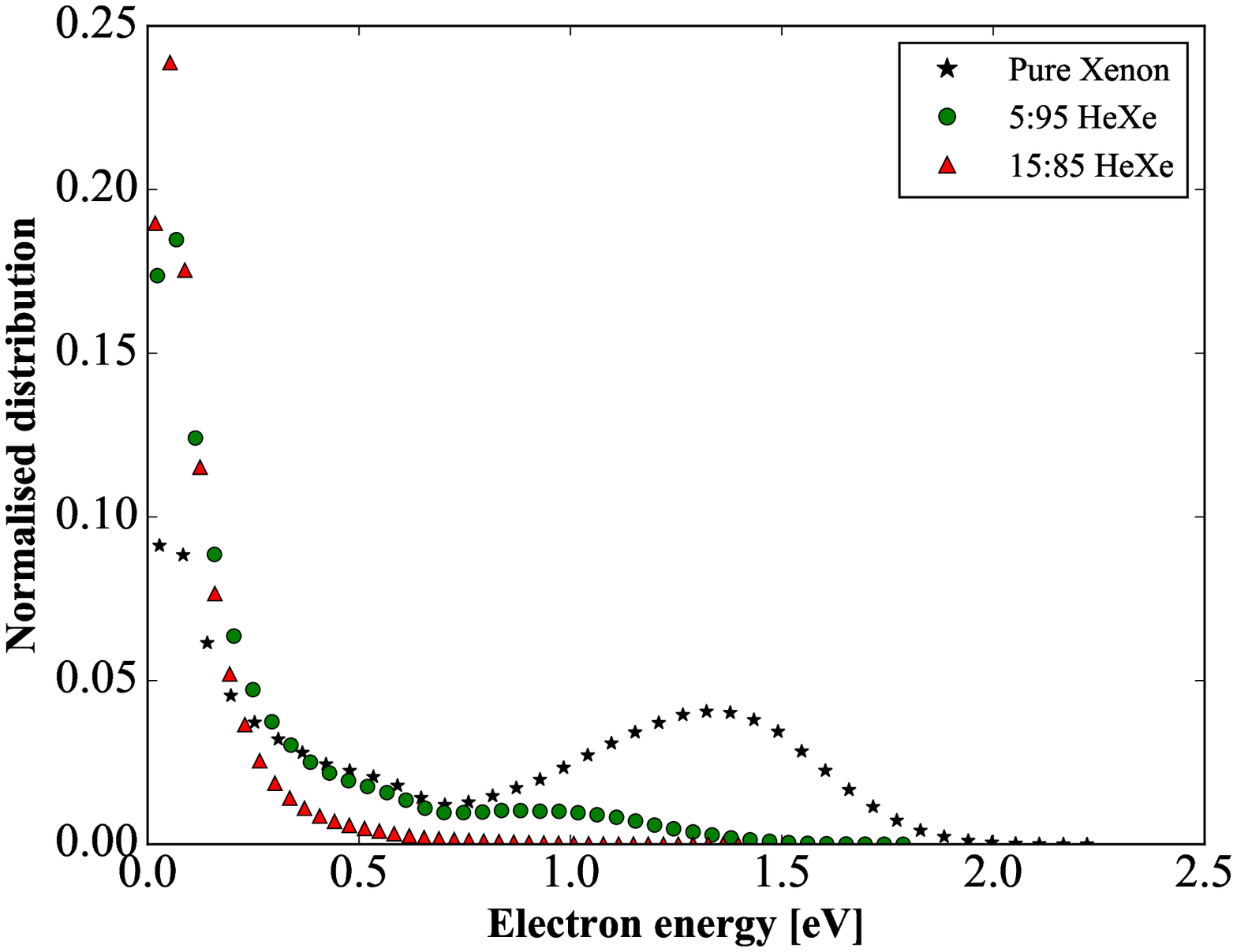}
	\caption{The \textit{top} figure shows electron-xenon and electron-helium cross section vs. the electron energy as extracted from \cite{Gas_xsec}. The Ramsauer minimum can be observed in the xenon cross section (solid black) while the helium cross section (dashed red) remains stable below 10~eV. The \textit{bottom} figure shows three energy distributions at 400~V/cm and 15~bar as computed by Magboltz: one in pure xenon (black) and the two others (green and red) in HeXe mixtures.}
	\label{fig:Xe_xsec}
\end{figure}

While drifting in a noble gas TPC, secondary electrons reach statistical equilibrium by balancing the energy gained through the action of the electric field with that lost in collisions with the environmental noble gas atoms.
The fact that electron energies (under the typical drift fields in TPCs) are far from the excitation level of the noble gas atoms implies that electron-atom collisions are elastic, allowing a fairly accurate estimate of the momentum transfer by using a classical kinematic calculation of two body collisions.
The momentum transfer efficiency depends, then, on the mass ratio of the two bodies. Assuming isotropic scattering, one can approximate the fractional energy loss averaged over all scattering angles by the formula:
\begin{equation}
	\label{eq_frac_loss}
	\frac{\delta \varepsilon}{\varepsilon}\sim\frac{2mM}{(m+M)^2}
\end{equation}
where m is the electron mass and M is the atomic mass of the noble gas. Table \ref{table_frac_energy} lists this value for all noble gases generated using eq. \ref{eq_frac_loss}. It must be said that elastic scatterings are not necessarily isotropic, but this assumption is reasonable for helium in the whole context of this paper \cite{Fon:1981}. As for xenon, this is a valid assumption for electron energies up to about 2.75~eV \cite{Fraser:1987fs}. As can be seen on the bottom panel of Fig. \ref{fig:Xe_xsec}, this condition is well fulfilled.

By contrast to this large increase of the momentum transfer for light gas, one could expect that the total energy loss of the electrons will remain approximately constant as the cross section at very low energies becomes much larger for xenon than for lighter atoms, by virtue of its larger size (`solid sphere' model). However, the existence of the Ramsauer minimum \cite{Kukolich:1968} in the xenon cross section that can be seen in Fig. \ref{fig:Xe_xsec} counteracts the increase in atom size, making the overall electron cooling of helium much more effective.
\begin{table}[h!]
	\centering
	\begin{tabular}{|l|c|r|}
		\hline
		He & $2.74\cdot10^{-4}$ \\
		\hline
		Ne & $5.44\cdot10^{-5}$ \\
		\hline
		Ar & $2.75\cdot10^{-5}$ \\
		\hline
		Kr & $1.31\cdot10^{-5}$ \\
		\hline
		Xe & $8.07\cdot10^{-6}$ \\
		\hline
	\end{tabular}
	\caption{Mean fractional energy loss of electrons in collisions against noble gas atoms.}
	\label{table_frac_energy}
\end{table}

Neon and helium are the two natural options with respect to table \ref{table_frac_energy}.
While neon is easier to manipulate, helium is much more promising in terms of performance, as can be expected from its higher cross section at eV energies.

\begin{figure}
	\includegraphics[width=90mm]{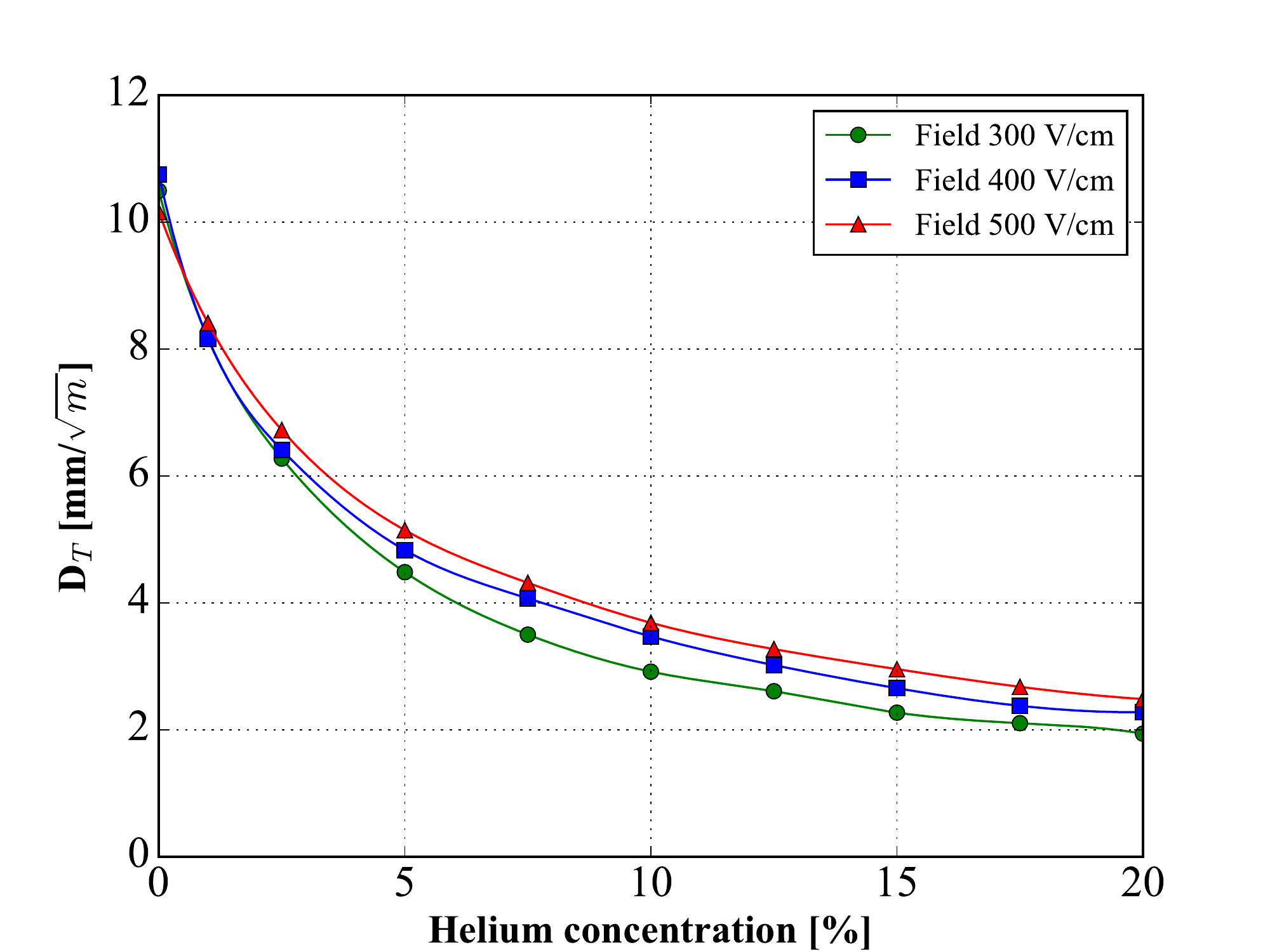}
	\caption{Transverse diffusion coefficient vs. the helium concentration in a HeXe mixture at 15~bar absolute pressure. Solid curves fitting the points are drawn to guide the eye of the reader.}
	\label{transverse_fig}
\end{figure}

We provide the results of simulations performed with the software Magboltz \cite{magboltz} shown in Fig. \ref{transverse_fig}.
The most relevant parameter to look at is the transverse diffusion, which is the dominating factor in the overall 3D diffusion.
Additionally, as we will show, the transverse diffusion component is the one that can be drastically reduced in the presence of additives.

These simulations were done assuming the standard working conditions in gaseous xenon-based \bbonu\ experiments, namely an operating pressure of 15~bar and an electric field ranging from 300~V/cm to 500~V/cm in the drift region. Unlike the longitudinal diffusion, the transverse diffusion is weakly affected by the electric field. The transverse diffusion coefficient is shown on Fig. \ref{transverse_fig} as a function of the helium concentration.

A transverse diffusion of 3.5~mm/$\sqrt{\textnormal{m}}$ is achievable with an admixture of 10\% of helium while 15\% of helium lowers it to the level of 2.5~mm/$\sqrt{\textnormal{m}}$ which is not far from the limit case of pure helium diffusion in these conditions : 1.8~mm/$\sqrt{\textnormal{m}}$. These values, which are less than twice higher than the thermal limit, remain a considerable improvement with respect to pure xenon.

These diffusion coefficients need to be considered in light of all instrumental effects in EL TPCs such as NEXT. Both the root-mean-square (RMS) spreads in the charge signals detected by the tracking plane photo-detectors, and the mean bias in position reconstruction, are relevant. The spatial distribution in the plane transverse to the drift direction, and for a point-like energy deposition in the gas, is affected by three factors:
\begin{itemize}
\item the transverse diffusion of the charge along drift, introducing the RMS transverse spreads quoted above;
\item the point spread function projected on the tracking plane due to isotropic light emission from a line segment at a fixed $x-y$ position within the EL region. A full Geant4 simulation, including reflected light, has been performed for the currently operating NEXT-White detector. In this case, for a 6~mm wide EL region and a 8~mm distance between the center of the EL region and the tracking plane, a 3.8~mm RMS transverse spread is obtained;
\item the sampling of the point-spread function at the tracking photo-detectors' positions. For a lattice of SiPMs at a 10~mm pitch in the $x-y$ plane, and conservatively assuming no light sharing between neighboring SiPMs, a 2.8~mm mean bias in transverse position reconstruction is obtained. This value should be taken as an upper limit on the position resolution induced by the SiPM pitch. The position bias is reduced by the effect of light sharing, and more elaborate algorithms can provide a much better transverse position estimate \cite{Simon:2017pck}.
\end{itemize}
As is apparent from the above numbers, and for a drift distance of one meter, a 10--15\% admixture of helium successfully reduces transverse diffusion effects to the same level of the RMS transverse spread introduced by the detector optics, and to the same level of the mean bias introduced by the SiPM pitch.

\subsection{Longitudinal diffusion}
\label{subsec:londiff}

The diffusion along the drift direction does not follow the same pattern as the transverse one.
Good descriptions of the longitudinal diffusion in the absence of inelastic collisions can be found in Parker and Lowke (1969) \cite{Parker:1969} and Skullerud (1969) \cite{Skullerud:1969}.
To summarize, the longitudinal diffusion is the summation of the purely thermal diffusion and an effect arising from the enhanced velocity along the drift field (`drift velocity').
\begin{figure}[b!]
	\includegraphics[width=0.48\textwidth]{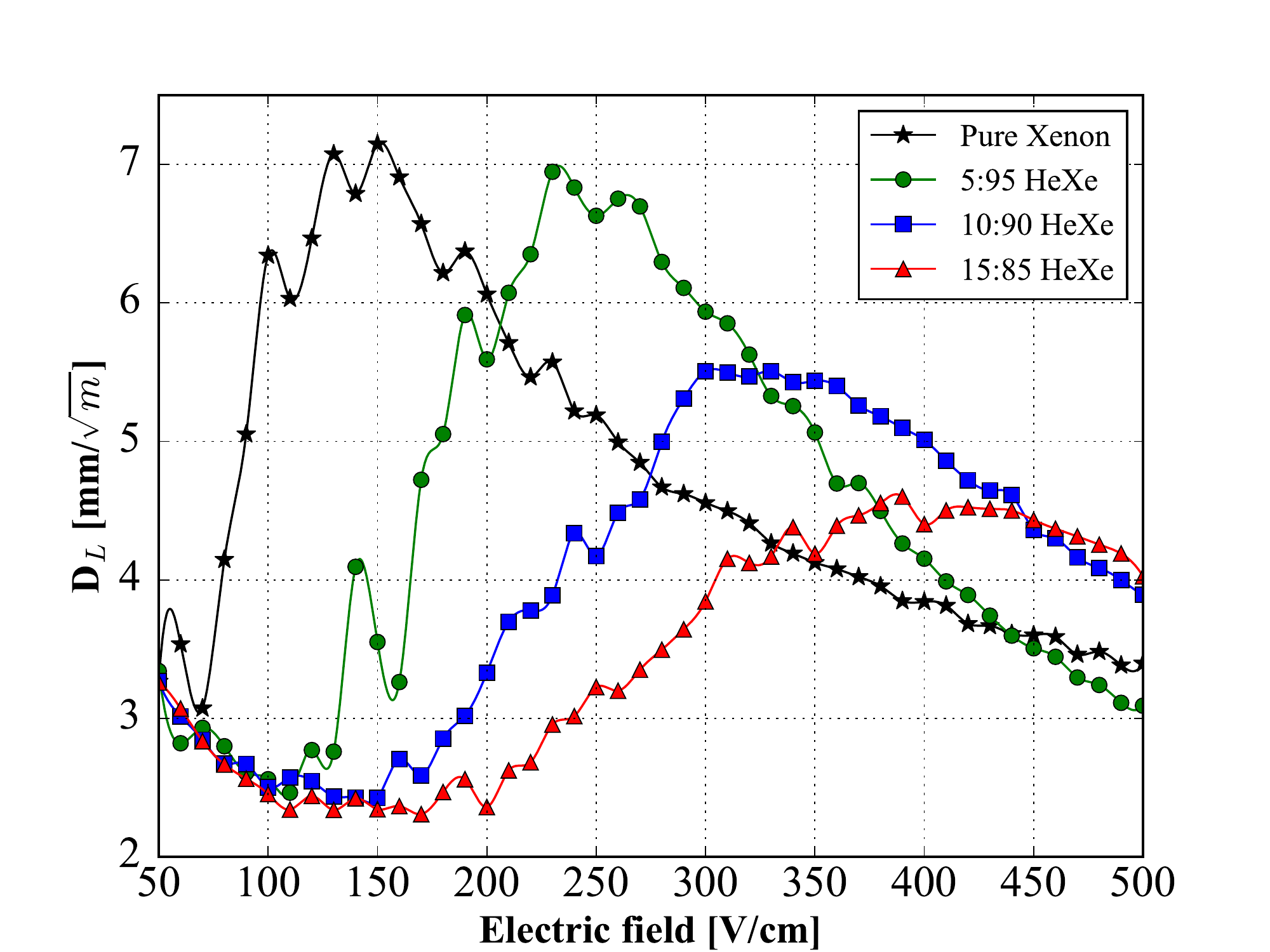}
	\caption{Longitudinal diffusion vs. the applied electric field for different mixtures. Solid curves fitting the points are drawn to guide the eye of the reader.}
	\label{longitudinal_fig}
\end{figure}
If the electron-atom collision frequency increases with energy then, while drifting, the electron swarm tends to spread along the drift direction due to thermal diffusion. Electrons in advance of the charge centroid will have an above average speed, which will rise their collision frequency hence reducing their instantaneous velocity due to momentum transfer.
Similarly, delayed electrons will experience fewer collisions leading to a greater instantaneous velocity along the drift.
These concurrent effects will narrow the electron cloud and effectively reduce the longitudinal diffusion.
In the case where the electron-atom collision frequency decreases with energy the effect described above is reversed.

When looking at the longitudinal diffusion of xenon-based mixtures one needs to remember that the elastic cross-section of xenon presents a minimum at 0.6~eV due to the Ramsauer effect (Fig. \ref{fig:Xe_xsec}).
In the drift region the energy distribution is located around that minimum which causes small changes to affect greatly the longitudinal diffusion due to the effect described previously.
This explains the observed `peak' shown in \cite{Pack:1992} when looking at the longitudinal diffusion against the reduced electric field or equivalently against the molecular admixture concentration as in \cite{Azevedo:2015eok}.
In the latter case these `Ramsauer induced peaks' appear at very low concentration, typically on the sub-percent scale, of the admixture due to their very strong cooling power.

In the case of helium-enriched admixtures this effect appears to be much broader than an actual peak as the Ramsauer minimum of xenon is not dominant (Fig. \ref{fig:Xe_xsec}). When looking at the longitudinal diffusion against the electric field for different admixture levels of helium in Fig. \ref{longitudinal_fig}, we can identify a ramping up region followed by a ramping down one.
For any given mixture, lowering the longitudinal diffusion requires the electric field to be high enough, around 500~V/cm. This is  specifically true for the 15:85 HeXe mixture whose `Ramsauer induced peak' reaches a maximum at 400~V/cm.
\begin{figure*}
	\includegraphics[width=90mm]{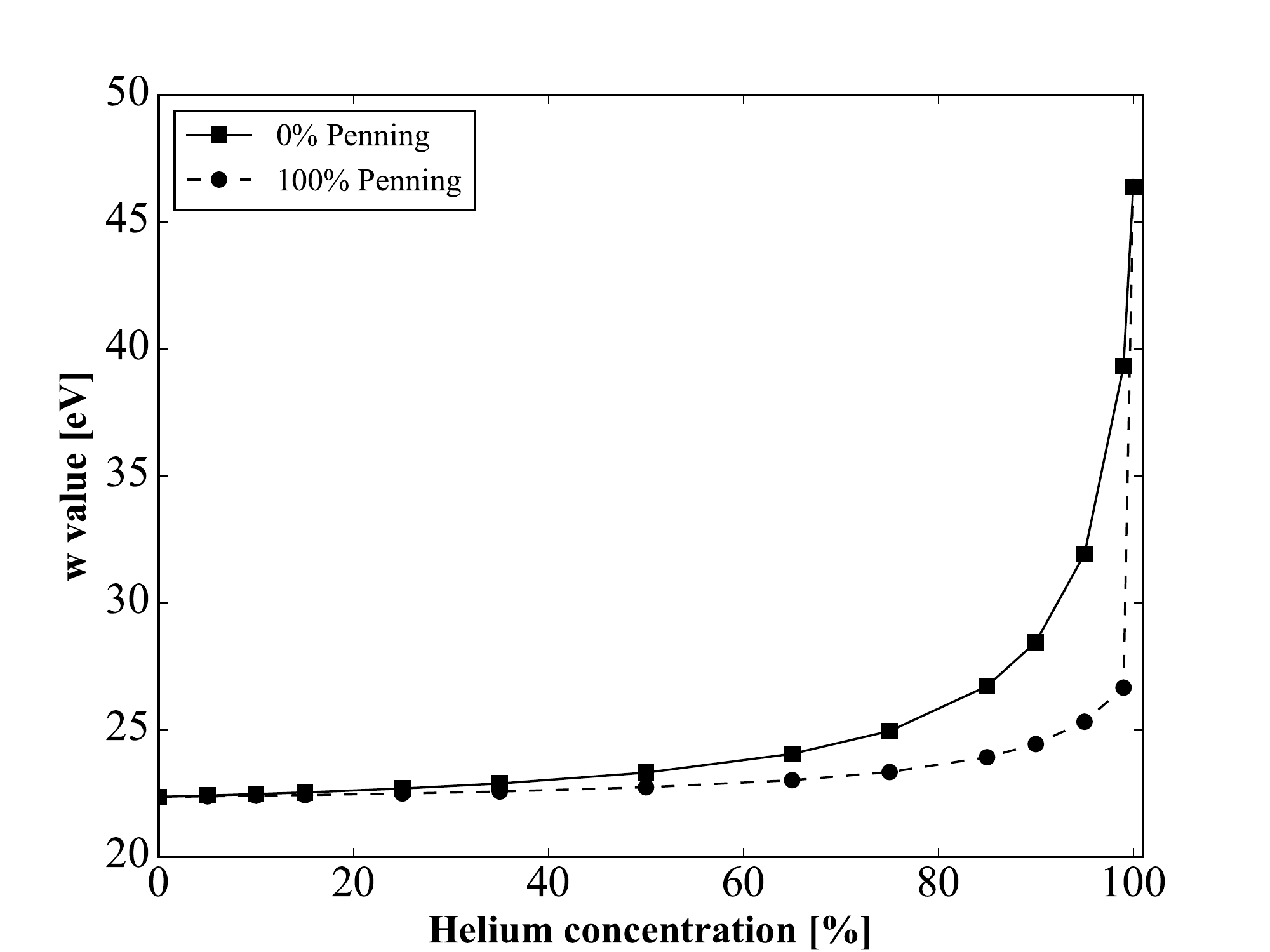}
	\includegraphics[width=90mm]{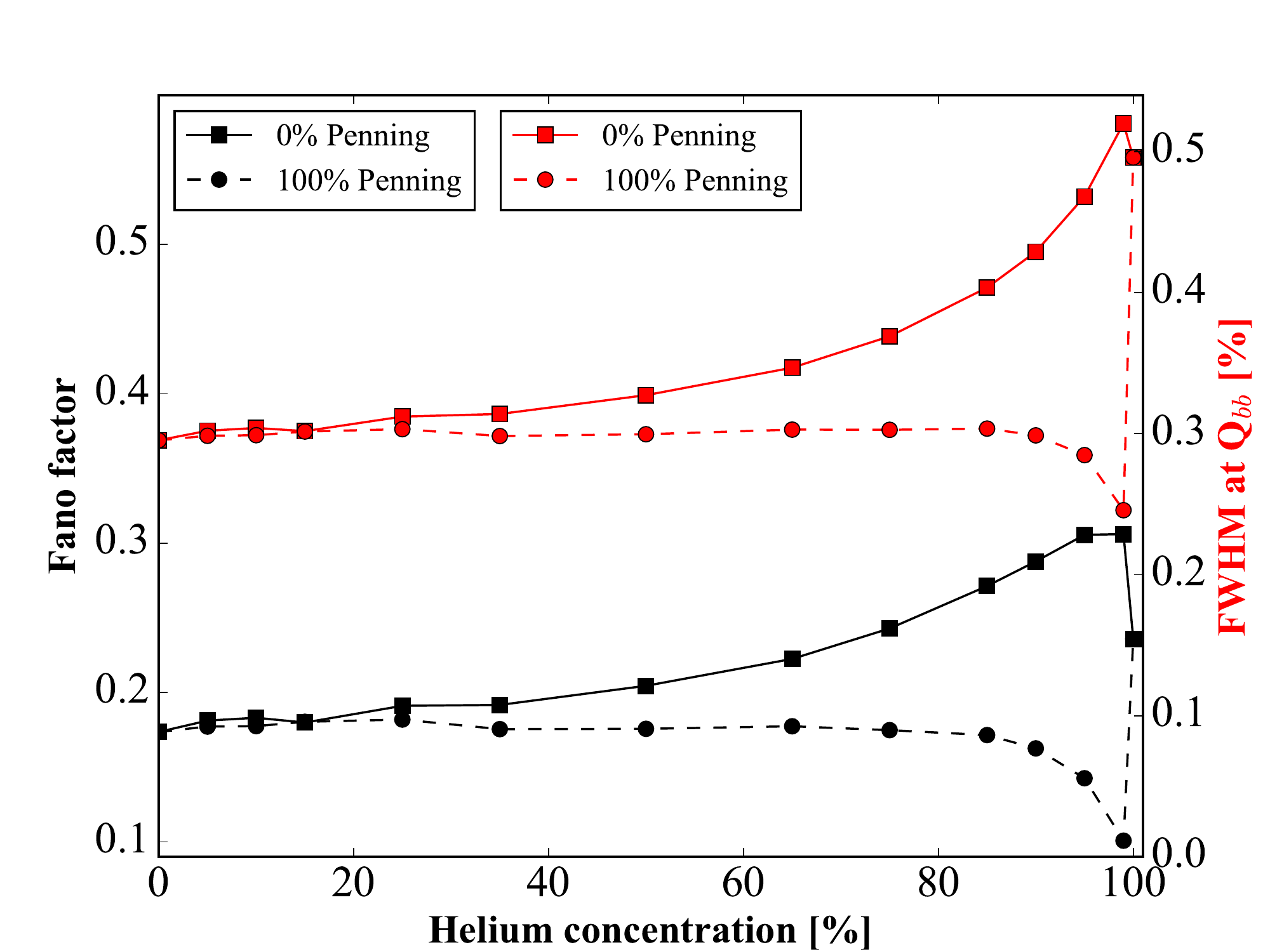}
	\caption{Average energy loss per ionization (\textit{left} figure) and Fano factor (\textit{right} figure) vs. the helium concentration in a HeXe mixture as computed by the software Degrad in a zero field scenario. The \textit{right} figure also includes in red the intrinsic energy resolution at Q$_{\beta\beta}$. Solid lines refer to the limiting case where there is no Penning effect while the dashed lines assume perfect efficiency in Penning transfer. In both cases the change remains minimal at low helium concentration, up to 20\%.}
	\label{fig:degrad}
\end{figure*}
It is to be noted that, at a glance, operating the 15:85 HeXe mixture at a field between 150 and 250~V/cm looks promising, but drawbacks in terms of charge recombination and finite electron lifetime are expected at these low fields. Whether or not it is interesting can only be decided by experimentation. But since earlier results from \cite{Bolotnikov:1999} are not encouraging in this respect, we concentrate here in the high field region. In this region, we can expect a longitudinal diffusion of the order of 4~mm/$\sqrt{\textnormal{m}}$ for a 10 and 15~\% concentration of helium.

As in the transverse case, the longitudinal diffusion coefficient needs to be considered in light of other instrumental effects. The spatial distribution along the drift direction, and for a point-like energy deposition in the gas, is affected by:
\begin{itemize}
\item the longitudinal diffusion of the charge along drift, introducing the RMS longitudinal spreads quoted above;
\item the spread introduced by the uniform light emission along the width of the EL region gap. A minimum width is required for high voltage considerations. For the 6~mm wide gap of the NEXT-White detector, a 1.7~mm RMS longitudinal spread is obtained;
\item the sampling of the photo-detector waveforms in the TPC time domain. For a 1~$\mu$s time sampling, a 1~mm/$\mu$s drift velocity, and conservatively assuming no light sharing among adjacent time samples in a waveform, a 0.3~mm mean bias in longitudinal position reconstruction is obtained.
\end{itemize}
Along this dimension, and given the assumption of a drift length of one meter, the tracking capability of the detector is therefore still dominated by the diffusion in the case of the helium-xenon mixtures considered.

\begin{figure}[ht!]
	\includegraphics[width=90mm]{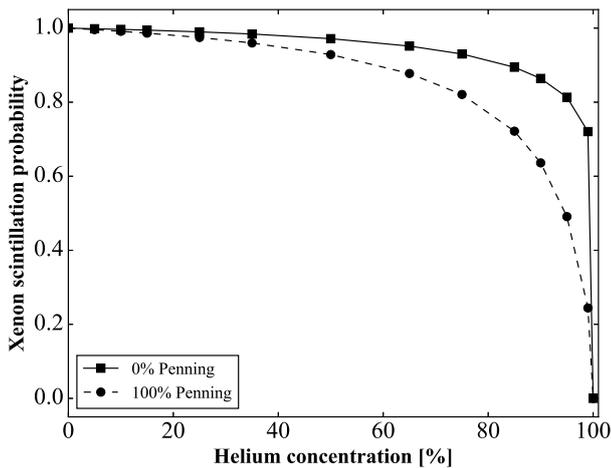}
	\caption{Scintillation probability of xenon vs. the helium concentration in a HeXe mixture. Solid lines refer to the limiting case where there is no Penning effect while the dashed lines assume perfect efficiency in Penning transfer.}
	\label{S1_proba}
\end{figure}

\section{Energy resolution}
\label{sec:eres}

The energy resolution is the main parameter of any \bbonu\ experiment as it reduces the size of the region of interest which in turn reduces proportionally the amount of background.
Also, the inherent background from the two-neutrino mode, that can only be rejected with a good energy resolution, limits the sensitivity of very large mass detectors  if resolutions at the percent level (FWHM) are not reached.
Hence the energy resolution at Q$_{\beta\beta}$ remains a fundamental performance parameter.
In the following section we study how the energy resolution is affected by the addition of helium to xenon.

\subsection{Fano factor and \textit{w} value}
\label{subsec:Fano}

The energy resolution is primarily affected by fluctuations in the production of electron-ion pairs.
The variance of the ionization is defined as:
\begin{equation}
	\label{fan_eq}
	\sigma_e^2 = F\bar{N}_e
\end{equation}
where F is the Fano factor and $\bar{N}_e$ is the average number of electrons produced at the energy E with $\bar{N}_e = E/\textit{w}$, \textit{w} being the average energy needed to produce one ionization.
The case of pure gaseous xenon has been studied extensively and is reported in the literature to have a Fano factor lying around 0.17 \cite{Henriques:2017rlj, Dias:1997x}, and a \textit{w} value of approximately 22~eV \cite{Dias:1997x}.
It is relevant to remark that the xenon Fano factor is very similar to the germanium Fano factor \cite{LOWE1997354}; the better energy resolution in germanium arises from the lower \textit{w} value and, consequently, from the higher number of total ionization electrons produced relative to xenon.
This intrinsic energy resolution is one of the great advantages of gaseous xenon in the search for \bbonu .

\begin{figure}[hb!]
	\includegraphics[width=90mm]{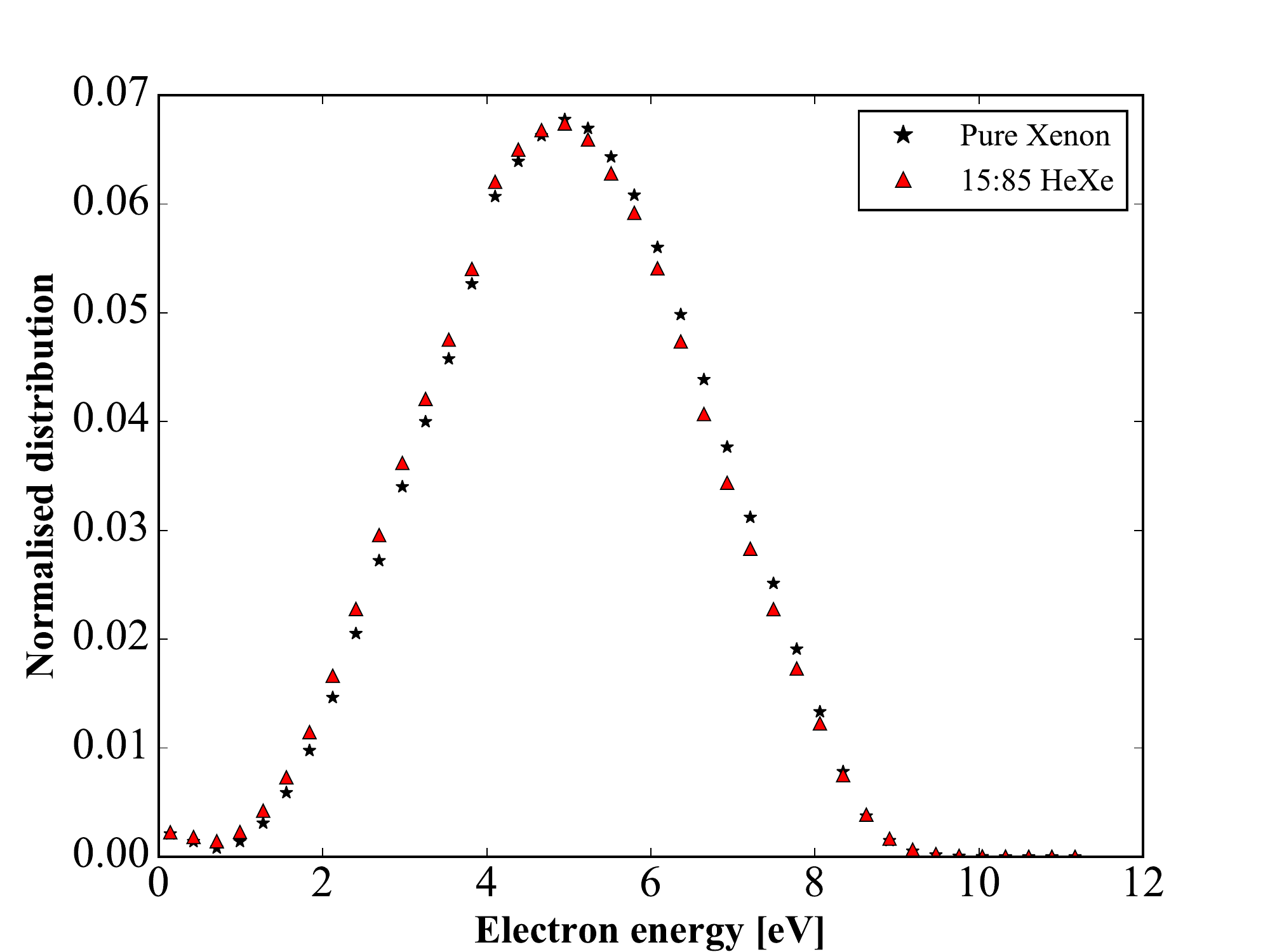}
	\caption{Normalised energy distributions as computed by Magboltz at a reduced electric field of 2.5~kV~cm$^{-1}$~bar$^{-1}$ in pure xenon and 15:85 HeXe. We can notice that there is very slight depopulation of the high energy tail of the distribution in the HeXe mixture.}
	\label{EL_distri}
\end{figure}

We computed those two parameters for the full absorption of single electrons of 2.48~MeV in different mixtures ranging from pure xenon to pure helium using the Monte-Carlo simulation software Degrad \cite{magboltz}. 
For event energies higher than the binding energy of the K-shell (30~keV in xenon), the Fano factor and the \textit{w} value do not change sizeably.
We assumed a gas at 15~bar and the simulation includes the Penning transfer (set at 100\%) ionizations occurring between helium and xenon as well as side effects such as Bremmstrahlung.
We enabled the Penning ionizations of xenon atoms by excited helium atoms but this topic deserves a specific discussion.

Penning transfer ionizations of xenon atoms by excited helium atoms were also allowed.
Helium excited states are all well above the ionization potential of xenon atoms and, among all the noble gases, the probability for a Penning transfer to occur at each collision is the highest in the case of helium-xenon collisions \cite{Bell:1968}.
When that is factored in with the fact that helium metastable states are very long lived (2$^3$S$_1$ lifetime is 131~min \cite{Hodgman:2009} and 2$^1$S$_0$ lifetime is 19.7~ms \cite{VanDyck:1971}) we can expect the Penning transfer probability to be very close to unity.
Whether or not all the energy is transferred from helium to xenon through ionization remains an open question as energy can as well be transferred through wavelength shifting by exciting a xenon atom.
To avoid any bias, and since no specific measurements have been made in the conditions discussed here, we simulated the two limit cases. First we considered no Penning ionization, i.e. all the energy goes to wavelength shifting and does not intervene in the charge carrier production. The second case is the one where all the helium excited states ionize xenon atoms. According to \cite{Sahin:2010ssz} the latter is probably closer to reality as hinted by their results for argon-xenon mixtures.
Results are displayed in Fig. \ref{fig:degrad}.

\begin{figure*}[ht!]
	\includegraphics[width=90mm]{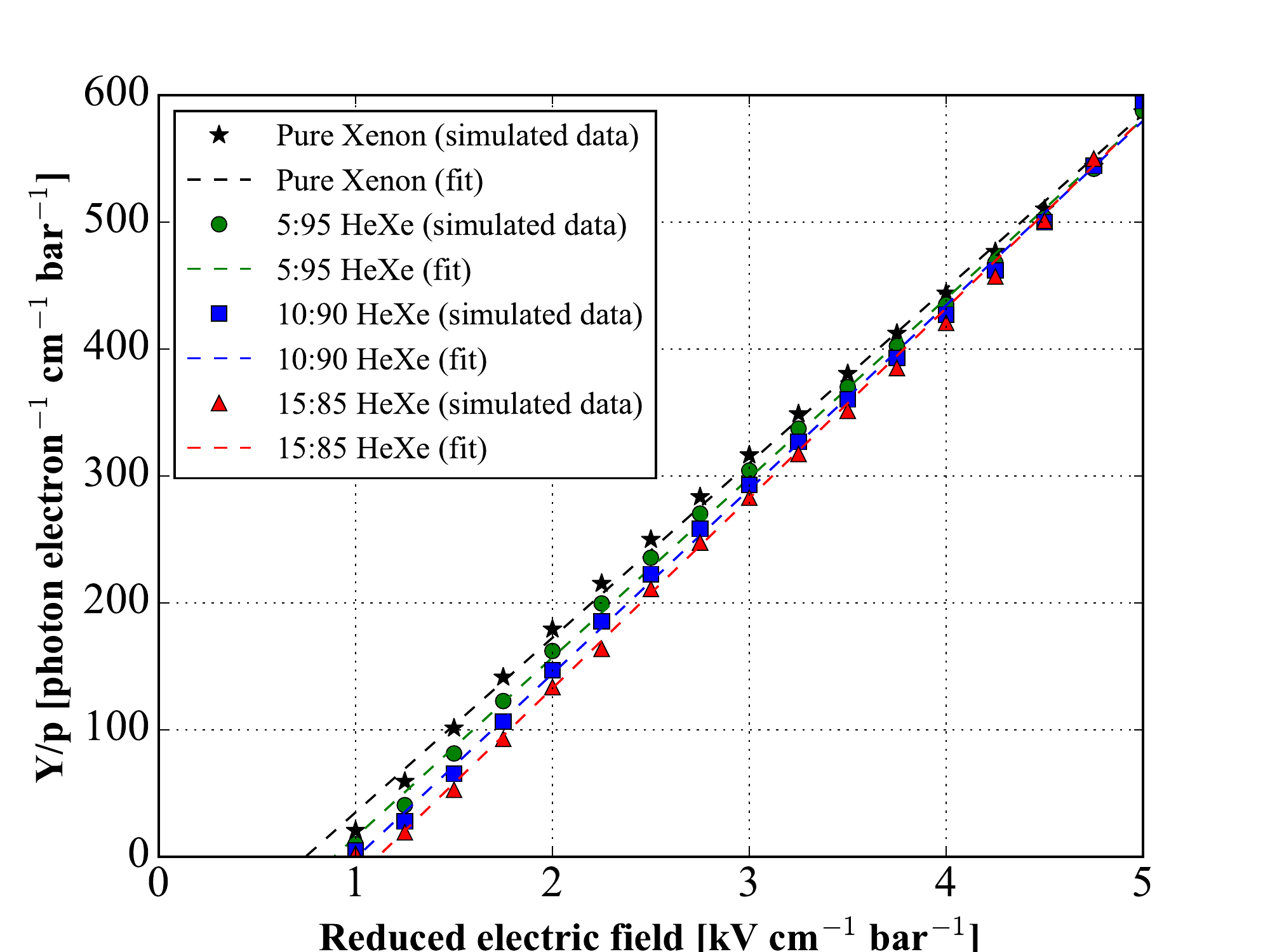}
	\includegraphics[width=90mm]{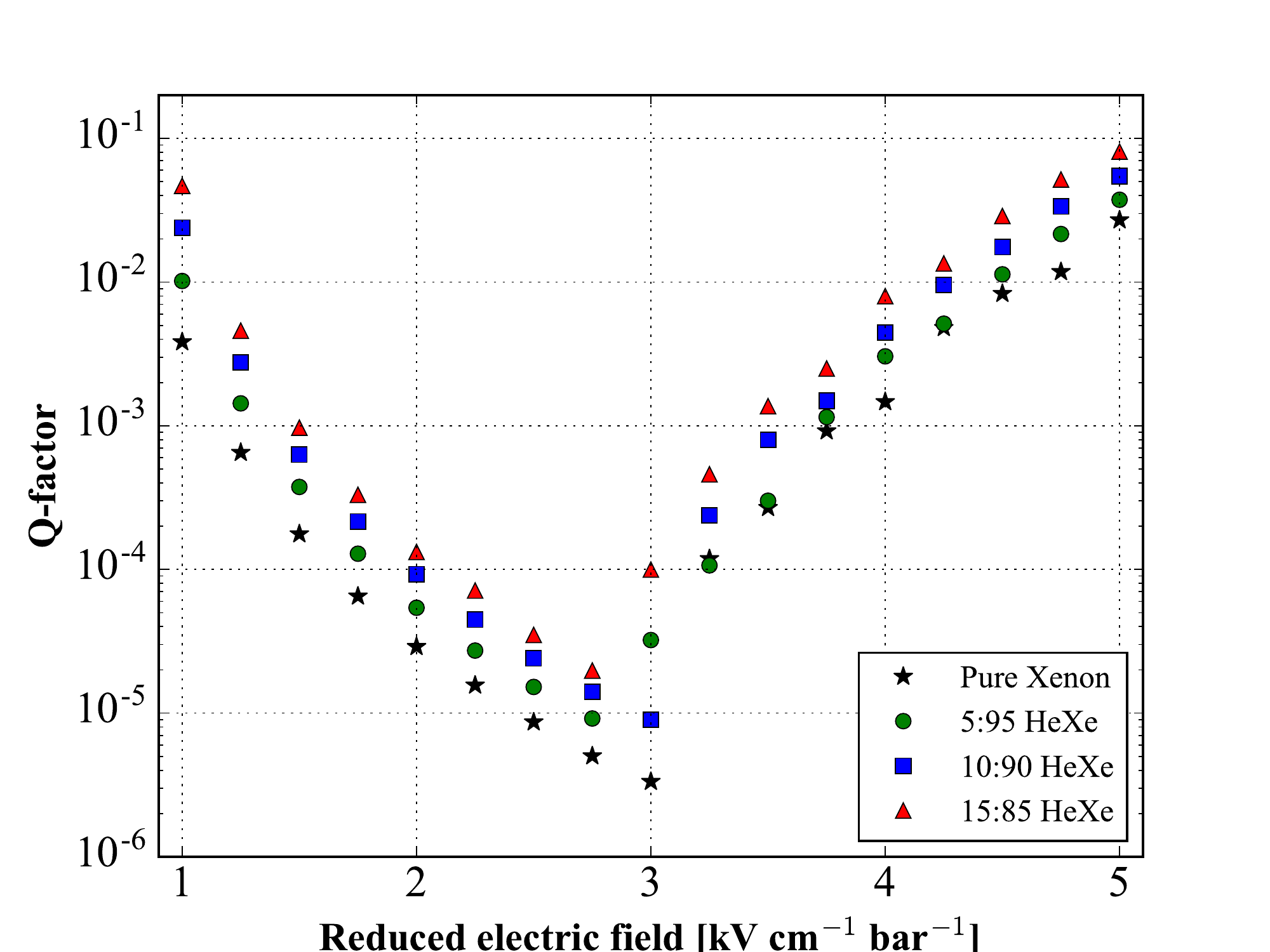}
	\caption{Reduced EL yield, Y/p, vs. the reduced electric field, E/p (\textit{left} figure). The points are obtained from simulation, while the lines are the corresponding linear fit. Q-factor vs. the reduced electric field (\textit{right} figure). The Q-factor is the quantitative definition of the yield fluctuation that we use (see eq. \ref{q_eq}).}
	\label{EL_fig}
\end{figure*}

There is no perceptible effect for the mixtures considered here, up to 20\% of helium, on both the Fano factor and the \textit{w} value.
The latter is limited by the one in pure helium, around 46~eV, but for the mixtures of interest its value is just marginally increased.
On the other hand, and for the same mixtures, the Fano factor is left unaffected. Consequently, the intrinsic energy resolution of helium-xenon mixtures remains very stable in helium-xenon mixtures as long as the helium concentration stays below 20\%.
Indeed, the Penning effect is unlikely to play any role, as the number of helium excited states is too low to make any difference (dashed and continuous lines in Fig. \ref{fig:degrad}).
It must be noted that these results show a trend similar to previous experimental and numerical studies on the evolution of the parameters in neon-xenon mixtures ranging from pure xenon to pure neon \cite{Dias:1999, Santos:2001}.

\subsection{Primary scintillation}
\label{subsec:s1}

Since Degrad provides detailed numbers of every excited state population we can briefly discuss how the primary scintillation behaves in helium-xenon mixtures, as detection of this prompt signal is necessary for a correct positioning of the events within the detector.
The scintillation properties of helium-xenon mixtures were studied in \cite{SAITO2007119} for relatively low partial pressures of xenon. This study shows that the scintillation intensity as well as the number of excited xenon atoms in the presence of an electric field saturates above 1 bar of xenon (for 10 and 0.657~bar of helium, 10\% xenon in the mixture), indicating that a detector with much more xenon than helium will have scintillation properties very close to a pure xenon detector with a small modification of the time distribution of the scintillation.

As in the previous section, the scintillation properties of helium-xenon mixtures were simulated for the two different assumptions aforementioned regarding the Penning transfer.
The scintillation probability with respect to pure xenon is shown in Fig. \ref{S1_proba}.
The pessimistic scenario in this case corresponds to the assumption of a 100\% Penning effect but, even in those conditions and as long as the helium concentration does not exceed 20\%, the scintillation level remains within 3\% of its value for pure xenon.
As expected, for the mixtures considered, the capability to detect the primary scintillation and to position the events in $z$ remains totally unaffected.

\subsection{Light yield}
\label{subsec:yield}

The reason behind using electroluminescence to amplify the ionization signal is that it provides a very low variance when compared to electron avalanches, and potentially allows Gas Proportional ElectroLuminescent Counters (GPELCs) to achieve an energy resolution fairly close to the intrinsic limit, given above.
When looking at the energy resolution versus the light yield in the EL stage it is apparent that a too low number of produced photons degrades the energy resolution. This effect is due to the high relative variance of the light production at low reduced field, but also due to electronics noise and finite photon statistics.
On the other hand, the energy resolution deteriorates at high field when the electrons get enough energy to ionize the gas atoms.
A regime of interest in terms of reduced electric field (E/P) for pure xenon has been identified in the range between 1.5 and 3.5~kV~cm$^{-1}$~bar$^{-1}$ in \cite{Oliveirabis:2011}. In this regime, the impact of the relative variance of the light production is very low, but higher than the fluctuations coming from the residual ionization, that is hence subdominant.

The first excited state of helium being 7.67~eV higher than the first ionization level of xenon, we expect helium not to play a direct role in the secondary light production of an EL TPC.
This is confirmed by looking at Fig. \ref{EL_distri}: unlike in the drift region, the normalised energy distributions in pure xenon and in HeXe mixtures keep the same shape albeit a very small quantitative difference in the high energy tail. In that energy region, interactions with xenon dominate again over those with helium (Fig. \ref{fig:Xe_xsec}), thus reducing the impact of the latter on electron cooling.
On top of that, a helium-xenon mixture remains totally transparent to the VUV xenon light.
Therefore, and keeping in mind the conclusion of the preceding section, we do not expect the total energy resolution of xenon to be hurt by the addition of helium. 

We used the scintillation model of \cite{Oliveira:2011xx} implemented through a simulation based on the Garfield++ toolkit \cite{garfield}. We studied the light yield for the 5:95, 10:90 and 15:85 HeXe mixtures as well as for pure xenon for comparison purposes.
The simulation generates electrons in a region that has a drift electric field of 400~V/cm across 0.5~mm before entering the EL stage. We took 5~mm as the length of the EL gap and 15~bar for the gas pressure, while the reduced electric field goes from 1.0 to 5.0~kV~cm$^{-1}$~bar$^{-1}$, with a step of 0.25~kV~cm$^{-1}$~bar$^{-1}$. The field maps have been produced with the COMSOL \cite{comsol} electrostatic module.
Each configuration consists of a sample of 10,000 initial electrons.

When looking at the results in Fig. \ref{EL_fig}, the first impression is that the HeXe mixtures stay very close to pure xenon in terms of light yield, which is great for energy resolution.
Everything else staying equal, we can notice a slight shift of the threshold toward higher fields the more we add helium. This is due to the residual cooling effect of helium, but this small shift can be overcome by increasing adequately the voltage across the EL gap.

To evaluate the fluctuation in the light production we used as a figure of merit the Q-factor, which is defined as follow:
\begin{equation}
	\label{q_eq}
	Q = \sigma_{EL}^2/N_{EL}^2
\end{equation}
The advantage of using the Q-factor is that it adds up directly with the Fano factor in the energy resolution formula, when restricted to the sole contributions of the intrinsic energy resolution and the electroluminescence process:
\begin{equation}
	\label{eq_eres}
	R_E = 2.35\sqrt{\frac{w}{Q_{\beta\beta}} [F+Q]}
\end{equation}
Also, as seen in the right plot in Fig. \ref{EL_fig}, the fluctuations in the light production increase the more helium is added.
But all in all this is not worrisome as these fluctuations are well below the Fano factor.
For the sake of the example we can take the specific case at 2.5~kV~cm$^{-1}$~bar$^{-1}$ and put together the results given by Degrad and Garfield++ in eq. \ref{eq_eres}: in pure xenon the \textit{w} and Fano factor given by Degrad are respectively 22.36~eV and 0.1736 while the Q-factor is 8.67$\cdot$10$^{-6}$ which leads to an energy resolution of 0.295\%; in 15:85 HeXe the Q-factor is 3.50$\cdot$10$^{-5}$, the \textit{w} is 22.43~eV and F is 0.1804 so we obtain an energy resolution of 0.302\%.
One can quickly remark, given the values, that the difference between both energy resolutions is due to the different \textit{w} value and Fano factor.

In reality, eq. \ref{eq_eres} contains a finite photon-statistics term, whose detailed evaluation is outside the scope of this work. But given that the scintillation yield (Fig. \ref{EL_fig}) remains largely unaltered, we do not expect the situation to worsen compared to a pure xenon experiment.
\section{Collateral advantages of helium as an additive}
\label{sec:colladv}

A helium-xenon mixture used in an EL TPC would provide a competitive diffusion while keeping the advantage of the pure xenon energy resolution. 
At the same time, and contrary to the general situation regarding molecular additives, light yields will be largely unaffected too (both primary and secondary).
But one can think of additional advantages over the use of either molecular additives or pure xenon.

\subsection{Drift velocity}
\label{subsec:vd}

\begin{figure}
	\includegraphics[width=90mm]{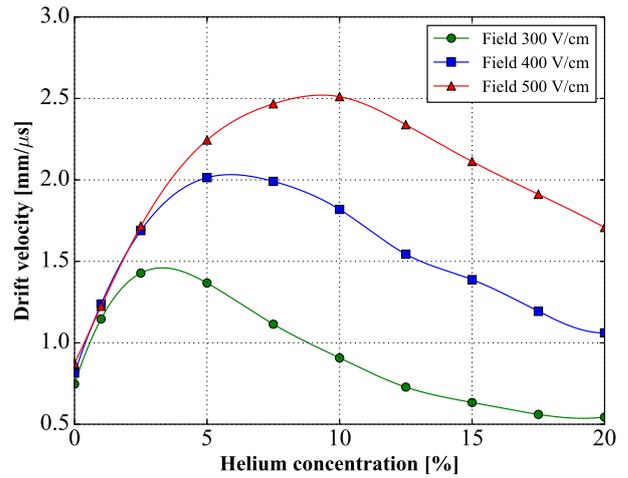}
	\caption{Drift velocity vs. the helium concentration in a HeXe mixture for three different fields. Solid curves fitting the points are drawn to guide the eye of the reader.}
	\label{vel_fig}
\end{figure}

In rare event search experiments, the drift velocity is not a crucial parameter.
According to Magboltz, whose simulation results can be seen in Fig. \ref{vel_fig}, a drift velocity more than twice the one in pure xenon can be achieved with helium-xenon admixtures.
In our favourite scenario of 15:85 HeXe at 400~V/cm the drift velocity is expected to be 1.7 times the one in pure xenon at the same field.
This would have a minor positive impact in the data acquisition process, allowing shorter data buffers and plausibly an increased lifetime at the same impurity concentrations.

\subsection{Technical blessings}
\label{subsec:tech}

In an EL TPC, gas purity is a crucial parameter because it plays a major role in the electron lifetime and light yield. 
Helium, being a noble gas and even more chemically inert than xenon, would be implemented quite easily in an existing pure xenon gas system, and no additional care would be required from the purification systems.
The relatively large amount of helium needed in the gas system makes it easy to monitor with commercial systems like an RGA.

As enriched xenon is both very rare and expensive, one can not afford to lose it.
The intended partial pressures of helium in the gas system, about 2~bar, would provide a very convenient way to hunt eventual leaks in the gas system.
In the case of micro-leaks helium is much more likely to escape, and in greater quantity than xenon, triggering a reaction from the system before losing any sizeable amount of xenon.

Also, the recovery of xenon is a technical requirement of any xenon-based neutrinoless double beta experiment.
Separating helium from xenon is very easy through cryogenic recovery because of the significant disparity between the boiling point of helium, 4.22~K, and the melting point of xenon, 161.4~K.
A few cycles of cryogenic recovery are enough to retrieve almost 100\% of the xenon.
\section{Helium and phototubes, a risky call}
\label{sec:optsens}

In the current state of the art, phototubes are the most commonly employed technology in measuring light levels similar to those produced by the primary and secondary scintillations detected in performing respectively the S1 trigger and the calorimetry.
And it is of public knowledge that helium atoms are a great danger for phototubes \cite{Incandela:1987dh}.
Indeed, phototubes require a very good vacuum level. Gaseous impurities cause afterpulses and lower the breakdown voltage of the phototubes. Once the internal pressure of helium is around $10^{-2}$ or $10^{-1}$~mbar at best it causes a total breakdown of the phototube, making it unusable.

The high permeability of glasses, especially fused silica, to helium is due to the fact that the holes in the amorphous structure of the glass are large enough to allow the small helium atoms to pass through them \cite{Norton:1953}.
Knowing the permeability constant of fused silica \cite{Norton:1953,Altemose:1961}, one can show using the empirical formula in \cite{Norton:1953} that it would take less than a couple of hours of operation at around 1 bar to fatally damage a phototube.

While this problem looks tough, the effort needed to overcome it may be worth it in view of the aformentioned discussions.
In the same studies on glass permeability to helium it is shown that the glass composition dictates its permeability.
Boro-silicate or soda-lime glasses are significantly less permeable to helium than fused silica.
A common interpretation is that the non glass-former elements like CaO, Na$_2$O or even PbO block the open areas in the glass structure which limit the number of open channels for helium atoms to diffuse, hence reducing the permeation constant.
While soda-lime glasses can handle an atmosphere of helium at 1~bar for as long as hundreds of days before seeing any afterpulsing, they are not a commercially available option for a radiopure experiment.

However, for decades, only an upper limit of helium diffusion through the crystal version of silica, crystalline quartz, could be set experimentally.
Some theoretical calculations \cite{Kalashnikov:2003} even demonstrate the incapability of helium atoms in their ground state to diffuse along the channels of crystalline quartz.

Operation within a helium atmosphere will require the use of optical windows that are able to withstand the gas pressure and prevent helium diffusion through them.
Synthetic sapphire is also a mineral commonly used as a window-material because of its good optical properties and mechanical robustness.
It has a much higher density, 3.98~kg/cm$^3$, than crystalline quartz, 2.65~kg/cm$^3$, which somewhat hints toward its non-permeability to helium.
In addition, the hexagonal compact structure of sapphire (Al$_2$O$_3$) leads the oxygen ions to almost achieve a perfect close packing of equal spheres \cite{Dobrovinskaya2009} maximizing the volume occupation of the crystal so, unlike crystalline quartz, there are no channels extending through the whole $c$-axis of sapphire.
This structure prevents helium atoms from moving through the crystal by forcing the atoms or ions to cross significant energy barriers. Such crossings are very unlikely to happen hence remaining unnoticeable at room temperature for reasonable periods of time.
A compilation of measurements of noble gas diffusion through minerals can be found in the review \cite{Baxter:2010}.

Since helium diffusion through sapphire has not yet been studied we can look at a very similar mineral: hematite.
Its crystal structure belongs to the hexagonal scalenohedral class of the trigonal crystal system which is also the case of sapphire (also known as corundum).
The extrapolation of the data measured in \cite{Lippolt:1993} sets the diffusion coefficient of hematite at room temperature below 10$^{-26}$~cm$^2$/s. This is about twenty orders of magnitude below the same value measured for glasses, typically ranging from 10$^{-8}$ to 10$^{-7}$~cm$^2$/s \cite{Swets:1961} at room temperature.
The activation energy in the temperature-dependent diffusion equation is the characteristic value describing the height of a potential barrier in the material, hence comparing those of a crystal and glasses is relevant to our discussion.
The hematite activation energy is 116~kJ/mol while the glass activation energies reported in \cite{Altemose:1961} range from 20~kJ/mol for fused silica to 52~kJ/mol in the case of an aluminosilicate glass.
Consequently, for our particular application, it seems as though we can assume that sapphire is unpermeable to helium.

\begin{figure}[ht!]
	\includegraphics[width=90mm]{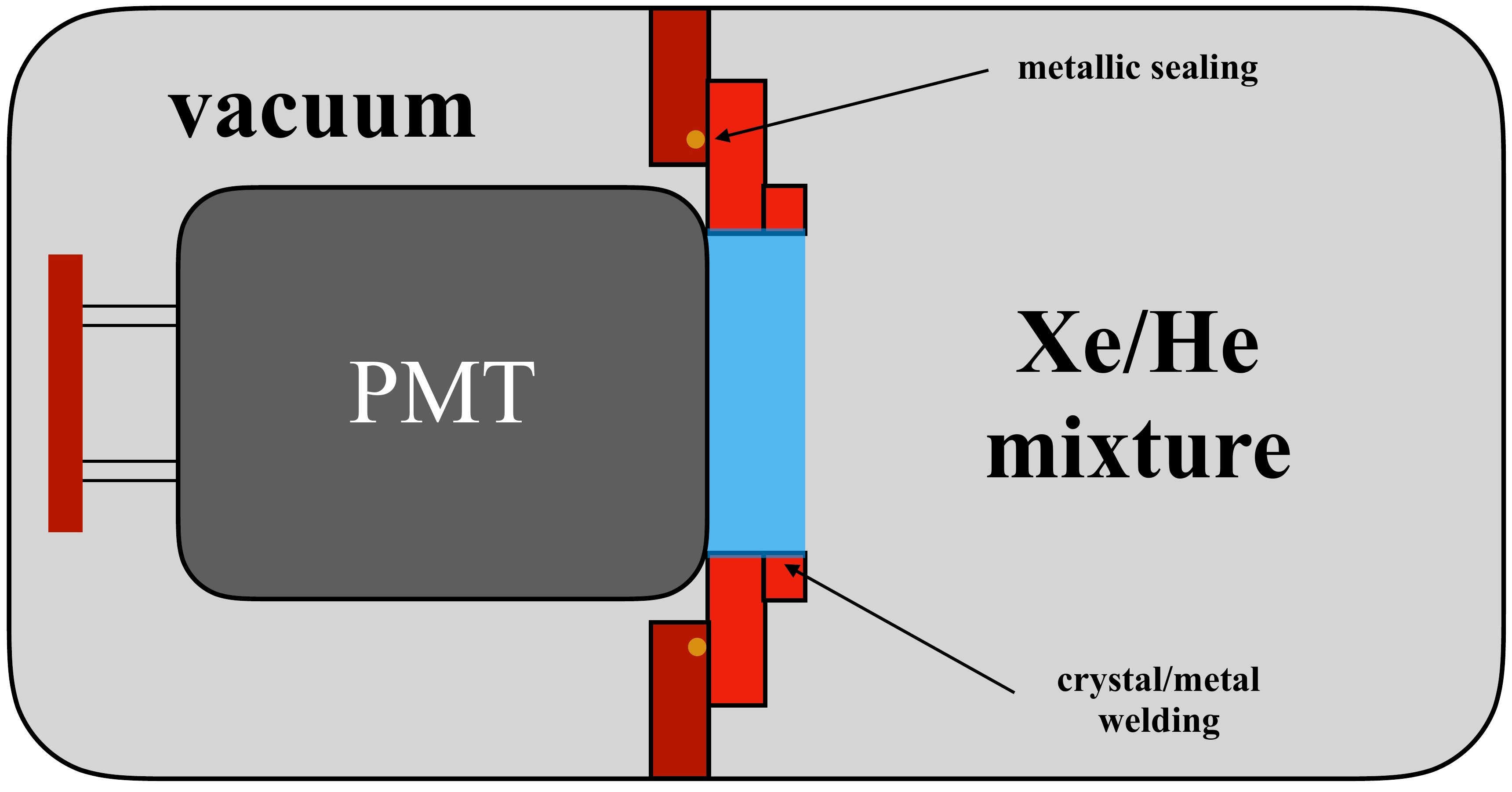}
	\includegraphics[width=90mm]{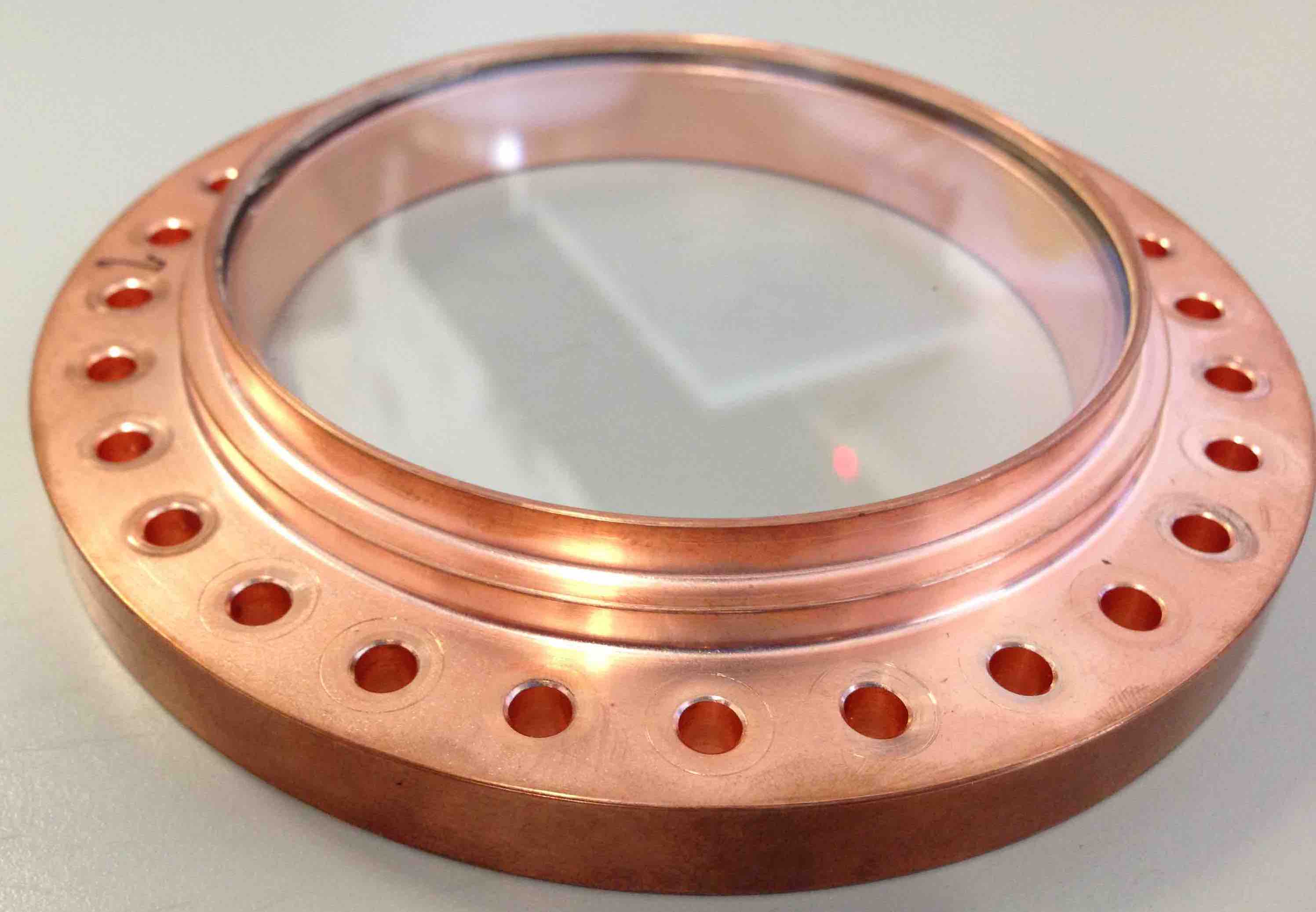}
	\caption{Scheme for the operation of PMTs in a helium-rich atmosphere (\textit{top} figure). The PMT itself remains at vacuum looking into the active volume through a crystalline quartz/sapphire window that can be coated with a wavelength shifter to maximize the light transmission. The crystal is welded to a metallic frame that can easily be mounted in a flange with a metallic sealing, avoiding helium diffusion into the PMT volume.
	The \textit{bottom} figure shows one of the NEXT-NEW PMT protective windows. A similar concept can be applied for operation within a helium atmsophere using a metallic sealing for this frame.}
	\label{fig:pmt_vac_press}
\end{figure}

One way to protect the PMTs is to encapsulate them in a vacuum vessel with a sapphire window separating them from the helium atmosphere. Sapphire optical properties offer a very good transmission of the blue light obtained from the VUV xenon light by a wavelength shifting coating (TPB).
Consequently it should be possible to operate an EL TPC with helium safely, either with crystalline quartz or sapphire. However, the welding of the crystal window remains a technical issue to consider.
Leak-proof sealants are well understood and metal-on-metal sealants should provide enough tightness to prevent any helium contamination of the PMTs. On top of that, one can maintain the PMTs inside a secondary vaccum to further prevent any damage. Fig. \ref{fig:pmt_vac_press} includes a schematic representation of this proposal as well as an example of a sapphire window used to protect PMTs.

A more drastic solution for the operation of helium-xenon detectors is to replace the PMTs by SiPMs. This solution will largely simplify the mechanical design associated with windows, weldings and operation of PMTs at vacuum. Current SiPMs sensors may offer a very interesting alternative if one can control the issues associated with the large number of channels and the capacitance increase associated with operating a large surface covered by SiPMs. Recent advances in this direction from different fields such as dark matter searches allow for an optimistic scenario, in which such issues are solved, to be considered.
\section{Conclusion and perspective}
\label{sec:conclusions}

We have proposed a new gas mixture that will enable a strong improvement on the topological signature of the current HPGXe technology by reducing the transversal diffusion by a large factor. On the other hand, the use of such mixture will reduce the amount of the source isotope in the detector. The final value of the helium concentration should be a compromise between an improvement of the background rejection factor and a reduction of the active mass that is needed to maximize sensitivity.
In this work we studied the impact of a helium admixture on the diffusion coefficients. The result of the simulations shows that a transverse diffusion of 2.5~mm/$\sqrt{\textnormal{m}}$ is achievable with 15\% of helium, improving by a factor of 4 the pure xenon value.
It must be noted that an earlier work described in \cite{Lanza:1987} on a helium-xenon scintillator was done in the context of medical imaging, but to the knowledge of the authors this work has not been further pursued.

On top of that, the intrinsic energy resolution remains unaltered, to within a few percent, in a helium-xenon mixture with respect to pure xenon. This is because the Fano factor, the number of ionization electrons and the optical properties of the gas do not change appreciably for those concentrations of helium. The photon yield in the EL region is slightly modified but not enough to become a limiting factor for the energy measurement.

The difficulty of operating phototubes close to a helium atmosphere also appears to be solved on paper. Windows made of good optical crystals such as crystalline quartz or sapphire can provide a helium-tight system that will allow for a safe operation of PMTs next to a helium atmosphere. On top of that, we encourage further tests using SiPMs as it will largely simplify the mechanical issues associated with PMTs.
A successful development of a helium-xenon gaseous optical TPC could have an impact in other research areas such as nuclear physics or dark matter searches.


 \section*{Acknowledgement}
The authors want to thank Igor Tolstikhin for providing useful references regarding helium diffusion through minerals.
The authors acknowledge support from the following agencies and institutions:
the University of Texas at Arlington; the European Research Council (ERC) under the Advanced Grant 339787-NEXT; the European Union's Framework Programme for Research and Innovation Horizon 2020 (2014-2020) under the Marie Sk\l{}odowska-Curie Grant Agreement No. 740055; the Ministerio de Econom\'ia y Competitividad of Spain under grants FIS2014-53371-C04; the Severo Ochoa Program SEV-2014-0398; DGD is supported by MINECO (Spain) under the Ramon y Cajal program (contract RYC-2015-18820) and Ryan Felkai under the Santiago Grisol\'ia program (contract GRISOLIA/2016/105).


\Urlmuskip=0mu plus 1mu\relax
\bibliography{bibli}

\end{document}